\documentclass[twocolumn,showpacs,aps,prl,amsmath,amssymb,floatfix,%
superscriptaddress]{revtex4}
\usepackage{graphicx}
\usepackage{dcolumn}
\usepackage{bm}
\draft

\begin{document}
\title{Kondo Lattice without Nozieres Exhaustion Effect}
\author{K. Kikoin}
\affiliation{Ben-Gurion University of the Negev, Beer-Sheva 84105,
Israel}
\author{M.N. Kiselev}
\affiliation{Physics Department, Arnold Sommerfeld Center for
Theoretical Physics and Center for Nano-Science,\\
Ludwig-Maximilians Universit\"at M\"unchen, 80333 M\"unchen,
Germany}
\date{\today}
\begin{abstract}
We discuss the properties of layered Anderson/Kondo lattices with
metallic electrons confined in 2D $xy$ planes and local spins in
insulating layers forming chains in $z$ direction. Each spin in
this model possesses its own 2D Kondo cloud, so that the Nozieres'
exhaustion problem does not occur. The excitation spectrum of the
model is gapless both in charge and spin sectors.  The  disordered
phases and possible experimental realizations of the model are
briefly discussed.
\end{abstract}
 \maketitle

Famous exhaustion problem formulated by Nozieres \cite{noz1} is a
stumbling stone on the way from exactly solvable Anderson or Kondo
impurity model \cite{WTAL} to the periodic 3D Anderson lattice
(AL) or Kondo lattice (KL) models, which are believed to provide
the generic Hamiltonians for mixed valence and heavy fermion
materials \cite{Hew,Rev}. The problem arises already for
concentrated Kondo alloys where the number or localized spins
$N_i$ is comparable with the number of sites $N=L^n$ in the
$n$-dimensional lattice. In this case the number of spin degrees
of freedom provided by conduction electrons in a KL is not enough
for screening $N_i$ localized spins. As an option a scenario of
dynamical screening was proposed, \cite{Colandr,Noz2} where only
part of spins screened by  Kondo clouds form magnetically inert
singlets. The low-temperature state of such KL is a quantum
liquid, where $N_{s}$ singlets are mixed with $N-N_{s}$ "bachelor"
spins, which hop around and exchange with singlets thereby
behaving as effective fermions. The Nozieres' exhaustion is
measured by a parameter $ p_N =N_i/\rho_0 T_K$ (the number of
spins per screening electron). Here $T_K \sim
\rho_0^{-1}\exp(-1/\rho_0 J)$ is the energy scale of Kondo effect,
$J$ is the exchange coupling constant in the single-impurity Kondo
Hamiltonian, $\rho_0$ is the density of states on the Fermi level
of metallic reservoir.

Second obstacle, which does not allow the extrapolation of Kondo
impurity scenario to KL is the indirect RKKY exchange $I_{jj'}$
between the localized spins, which arises in the 2nd order in $J$
or in the 4th order in $V$ (hybridization parameter in the generic
AL Hamiltonian). The corresponding energy scale is
\begin{equation}\label{coupc}
\rho_0 I = \rho_0J^2\chi_{jj'}^c \sim (\rho_0J)^2,
\end{equation}
where $ \chi_{jj'}^c=N^{-1} \sum_{{\bf q}}\chi_c({\bf q})\exp
(i{\bf q}\cdot{\bf R}_{jj'}) $ and $\chi_c({\bf q})$ is the spin
susceptibility of the electron gas. The Fourier transform of
$\chi_c({\bf q})$ is an oscillating function, which strongly
depends on the distance $R_{jj'}$. If $I<0$ at an average
inter-impurity distance, and $|I| \sim T_K$, then the trend to
inter-site antiferromagnetic (AF) coupling competes with the trend
to
 the one-site Kondo singlet formation (Doniach's
dichotomy \cite{doni}).

This competition prompted several possibilities to bypass the
exhaustion limitations. According to a scenario offered in
\cite{KKM,kikiop}, in the critical region $|I| \sim T_K$ of
Doniach's phase diagram, where the AF correlations are nearly
suppressed by the on-site Kondo coupling, the spin liquid phase
enters the game. This phase is characterized by the energy scale
\begin{equation}\label{coups}
\rho_0{\cal I}(T) = \rho_0 J^2 \chi^s_{jj'}(T)
\end{equation}
where $\chi_{jj'}^s $ is the spinon susceptibility. The condition
${\cal J}(T_K)$$>$$T_K$ is easily achieved both in 3D and 2D case.
The Kondo screening is then quenched in the weak coupling regime
at $T$$>$$T_K$, so that the spin degrees of freedom remain
decoupled from the electron Fermi-liquid excitations both at high
temperatures $T$$\gg$$T_K$  and at low temperatures $T$$\ll$$T_K$
(Curie and Pauli limit for magnetic response, respectively). At
$T$$\to$$0$ the KL behaves as a two-component Fermi liquid with
strongly interacting charged electrons and neutral spinons
\cite{kik}. This scenario develops on the background of strong AF
correlation. It includes the possibility of ordered magnetic
 phases with nearly screened magnetic moments and, in particular,
the quantum phase transitions. Due to separation of spin and
electron degrees of freedom, the Luttinger's theorem in its
conventional Fermi-liquid form is invalid in this state:
$f$-electrons represented by their spin degrees of freedom give no
contribution in formation of the electron Fermi surface. Such
state is referred as a "small Fermi surface regime" in current
literature.

Another scenario for small Fermi surface regime was proposed in
\cite{Sent1}. This scenario appeals to systems where the magnetic
order is either fragile or entirely absent due to magnetic
frustrations (e.g., to triangular lattices). A spinon gap carrying
unit flux of $Z_2$ gauge field is expected to arise in spin
subsystem, and this gap prevents formation of Kondo singlets for a
finite range of $T_K$. As a result the Nozieres' exhaustion does
not occur, and fractionalization of excitations into spin-fermions
and electrons exists like in the previous case.  A possibility of
forming the spin liquid with $U(1)$ gauge group and spin density
wave ground state has also been pointed out in \cite{Sent1}.

In the present paper we propose another paradigm for fermion
fractionalization in Kondo lattices, which possesses the generic
properties of KL but is not subject to the exhaustion limitations.
This paradigm may be realized in strongly anisotropic Kondo
lattices, where the metallic electrons are confined in 2D planes
interlaid by insulating layers containing magnetic ions. Then the
3D reservoir of screening electrons is defragmented into $L$
planar reservoirs. Each plane still possesses the macroscopic
number of spin degrees of freedom $\sim L^2$ enough for Kondo
screening, {\it provided} the concentration of magnetic centers
per metallic plane remains small. The spin liquid features may be
observed in these  systems if the distribution of magnetic centers
is also anisotropic, namely, if they form chains oriented in $z$
direction, and the inter-chain interaction is negligibly small.
\begin{figure}
\includegraphics[width=0.18\textwidth]{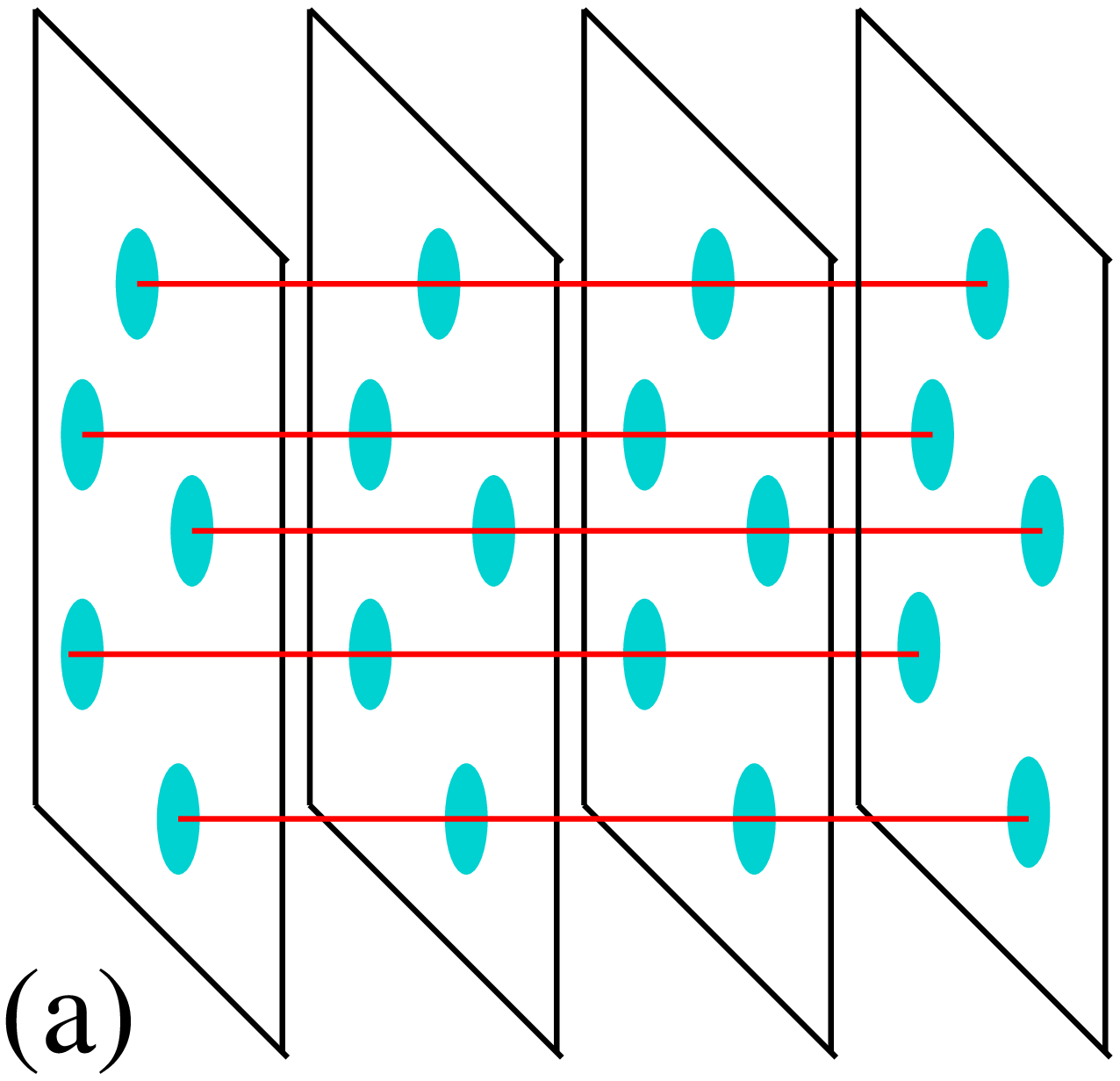}
\;\;\;\;\;\;\;
\includegraphics[width=0.22\textwidth]{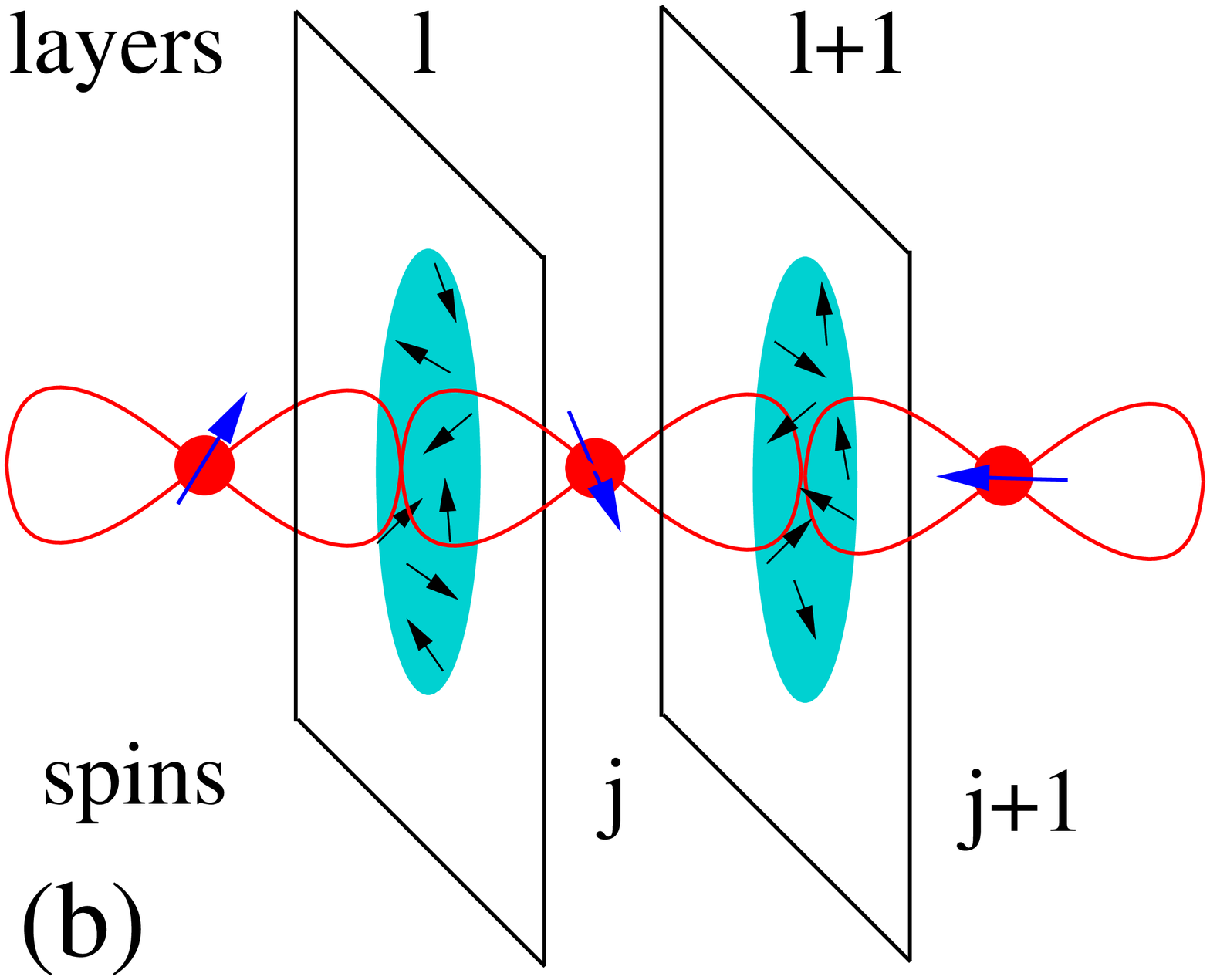}
\caption{\label{fig:f1a} (a) Layered lattice of spatially
separated charges in planes and spins in chains. (b) A fragment of
a chain with Kondo clouds formed as "shadows" in metallic layers.}
\end{figure}

Leaving the discussion of experimental realization of such systems
for concluding section, we begin with the theoretical description
of an ideal configuration, where all chains penetrate the stack in
$z$ direction (Fig.\ref{fig:f1a}a). The AL Hamiltonian for
 the quasiperiodic model of
conduction electrons confined in metallic layers ($xy$ plane), and
magnetic ions  localized in insulating layers between metallic
planes is
\begin{eqnarray}
H&=&\sum_{l{\bf k}\sigma}\epsilon_{k}c^\dagger_{l{\bf
k}\sigma}c_{l{\bf k}\sigma}+\sum_{j\sigma}\left(\epsilon_d
n_{j\sigma}^d
+\frac{1}{2}U n^d_{j\sigma}n^d_{j\bar\sigma}\right)\nonumber\\
&+&V\sum_{jl}\sum_{{\bf k}\sigma}\left(c^\dagger_{l{\bf
k}\sigma}(d_{j\sigma}+ d_{j+1,\sigma})+h.c\right) \label{h1}
\end{eqnarray}
Here ${\bf k}$ is a 2D wave vector, the discrete indices numerate
metallic layers $l$ with a lattice constant $a_\|$ and magnetic
sites $j$ along the chains with a spacing $a_z.$ The coupling
constant $V$ characterizes hybridization between itinerant 2D
electrons in a plane $l$ and localized states in two adjacent
chains $j$, and $j+1$ (Fig.1b). We treat the electrons in metallic
planes in terms of Bloch waves $c_{l{\bf k}\sigma}$, while the
localized electrons are characterized by Wannier functions
$d_{j\sigma}$. The lattice is quasiperiodic in a sense that the
periodicity of magnetic sites in the $xy$ plane is not demanded,
but  the average distance $\lambda$ between the impurities within
a layer exceeds the radius of Kondo cloud, i.e. satisfies the
condition $\lambda \ll \hbar v_F/T_K$ ($v_F$ and $T_K$ are Fermi
velocity of 2D electrons and Kondo temperature, respectively).
There is no interaction between the chains under this condition,
and a single chain represents the $z$ component of excitation
spectrum. On the other hand, all chains contribute to the
$xy$-component of the spin and charge response of the AL. The
effects associated with the inter-chain exchange will be discussed
in the concluding part.

We came to a situation where $L$ two-dimensional Fermi reservoirs,
each with capacity $L^2$, screen $N_i$ magnetic moments arranged
in such a way that the effective  concentration of these moments
per metallic layer is $n_i=N_i/L^2$ satisfies a condition
$n_ia_\|^2 \ll 1$. This capacity is enough to form screening Kondo
cloud for each magnetic site within a given layer $l$
independently of all other sites belonging to the same layer. On
the other hand, two magnetic ions localized one above another in
neighboring insulating layers $j$, $j+1$ share the same metallic
screen (see Fig. 1b). Replicating these dimers along $z$ axis, one
we come to a system of spin chains, interacting with a system of
metallic layers stacked up in the $xy$ plane. Elimination of the
hybridization term $V$ in the Hamiltonian (\ref{h1}) in accordance
with the standard Schrieffer-Wolff procedure, results in effective
exchange Hamiltonian for each chain,
\begin{equation}
H_{int}^{cd}=\sum_{j=1,k,k'}^{N_i}J_{\bf kk'} \vec{s}_{l+1,{\bf
kk'}} \left(\vec{S}_{j}+\vec{S}_{j+1}\right) \label{h2}
\end{equation}
(see Fig. \ref{fig:f1a}b).  Here~ $\vec{s}_{l+1,{\bf kk'}} =
\frac{1}{2}$$c^\dagger_{l+1,{\bf
k}\sigma}$$\vec{\sigma}_{\sigma\sigma'}$$c_{l+1,{\bf k'}\sigma'}$,
$\vec{S}_{j}=\frac{1}{2}$$d^\dagger_{j,\sigma}$$\vec{\sigma}_{\sigma\sigma'}$$d_{j,\sigma'},$
 and an exchange integral is estimated as $J\sim
V_{\bf k}^*V_{\bf k'}/U$.

Thus, the original model is reduced to the anisotropic KL formed
by the system of 1D spin chains penetrating the stack of 2D
metallic layers. Each spin creates two Kondo clouds in adjacent
planes, and two neighboring spins see each other through a
metallic screen by means of indirect RKKY-like exchange. This
exchange may be either ferromagnetic or antiferromagnetic. In this
work the latter case is considered in terms of the Doniach's
dichotomy \cite{doni}. In our model this dichotomy should be
reformulated. Since the long-range AF ordering is impossible in 1D
chain, two competing phases are Kondo singlet and spin liquid.
Complete Kondo screening is not forbidden by Nozieres' exhaustion
principle, since the 2D screening layer is available for each spin
in the chain. The Kondo screening is characterized by the energy
scale $T_K.$ Thus,  the competing phases in the anisotropic KL are
the Kondo singlet phase and the homogeneous spin liquid of RVB
type with the energy scale given by Eq. (\ref{coups}).

In order to describe the Doniach-like phase diagram we adopt the
method of \cite{kikiop}. Namely, we derive an effective action
functional by integrating out all "fast" fermionic degrees of
freedom with the energies $\sim D_0$, where $2D_0$ is the
conduction bandwidth. The "slow" modes give us a hydrodynamic
action. Due to strong quasi 1D anisotropy there is no need in
appealing to the mean-field approximation. After elimination of
conduction electrons with the energies $D_0
> \varepsilon
> T$ in metallic layers, the coupling $J$ is enhanced, $J \to \tilde J$$ =
1/\rho_0\ln T/T_K$. The indirect RKKY-like spin-spin interaction
mediated by the in-plane electrons arises along the chains:
\begin{equation}\label{h3}
H^{dd}_{int}= -I \sum_{j,\sigma\sigma'}
d^\dagger_{j\sigma}d_{j+1,\sigma}d^\dagger_{j+1,\sigma'}d_{j\sigma'}.
\end{equation}
Here $I$ is  defined in Eq. (\ref{coupc}) with $
\chi_{j,j+1}^c=N^{-1} \sum_{{\bf q}}\chi_c({\bf q})\exp
i(q_za_z).$ Since $V/U \ll 1$, we adopt the nearest-neighbor
approximation for RKKY interaction.

Up to this moment we treated the spin chains in a single-site
approximation. This approximation is legitimate until $T\gg T_K
\sim I$. To move further, we decouple the Euclidean action of the
model (\ref{h2}), (\ref{h3})
\begin{equation}
{\cal A}=\int_0^{\beta} d\tau \left[\sum_j\left(\bar c {\cal
G}_0^{-1} c + \bar d {\cal D}_0^{-1}d\right)
-H_{int}^{cd}-H_{int}^{dd}\right] \label{a1}
\end{equation}
by means of the Hubbard-Stratonovich scheme \cite{ren} in terms of
the fields $\Delta_{j,j\pm 1}\to
\sum_\sigma\left(d^\dagger_{j\sigma}d_{l\pm 1,\sigma}+c.c\right),$
$ \phi_{l}\to \sum_{\bf k\sigma}\left(c^\dagger_{l-1,{\bf
k}\sigma}(d_{j,\sigma}+d_{j+1,\sigma})+c.c\right)$
\cite{lark,kikiop}. Here ${\cal G}_0^{-1}=\partial_\tau
-\epsilon(-i\nabla)+\mu$ and ${\cal D}_{loc}^{-1}=\partial_\tau
-i\pi/(2\beta)$ are bare inverse single particle Green's Functions
(GF) for conduction electrons and local spins, respectively,
$\beta=1/T$. The field $\phi$ describes the single-site Kondo
screening and the field $\Delta$ stands for the spinon propagation
along the 1D spin chain with AFM coupling. The single occupancy
constraint $d^\dagger_{j\uparrow}
d_{j\uparrow}$$+$$d^\dagger_{j\downarrow} d_{j\downarrow}$$=$$1$
is preserved at each site in the chain  by the semi-fermionic
transformation \cite{pop}. These two fields resolve the Doniach's
dichotomy, because the long-range AF order is absent in 1D.

 We appeal to the uniform resonance valence
bond (RVB) spin liquid state \cite{lark} and treat the spinon
modes as fluctuations around the homogeneous solution in a $nn$-
approximation, $\Delta_{j,j\pm 1} \to =\Delta^u_{j,j\pm 1} -
\bar\Delta$ with
    $$\bar\Delta^2(\beta)= \beta^{-1}\int_0^\beta \Delta(\tau)\Delta(-\tau)d\tau
    .$$
For this sake we add and subtract $\bar \Delta$ in the inverse GF.
The non-local inverse spinon GF ${\cal D}^{-1}=\partial_\tau -
\Delta_{j,j\pm1}-i\pi T/2.$ has to be expanded in terms of
$\Delta-\bar \Delta$. Now the two interacting components of
bose-like modes in two-sublattice chain are spinons and Kondo
clouds represented in effective action by $\Delta_{j,
j+1}\Delta_{j+1,j}$ and $\phi_{l, l+m}\phi_{l+m,l}~ (m=0,1)$,
respectively. The charged $\phi$-mode acquires dispersion due to
the non-locality of
 $\tilde J_{lj}$, while the in-plane dispersion of
conduction electrons in Kondo clouds is integrated out. The
neutral spinon mode is dispersive by its origin. The action in
these terms is
\begin{eqnarray}
&&{\cal
A}_{eff}=\sum_{jl,\omega_n}\left(\frac{|\phi_{l,l}(\omega_n)|^2}{\tilde
J_{lj}}+\frac{|\Delta_{j,j+1}(\omega_n)|^2}{I}\right)
+\;\;\;\;\;\;\;\;\;\label{a2}\\
&&Tr\log({\cal G}_0^{-1})+Tr\log\left({\cal
D}^{-1}(\Delta_{j,j+1})+{\cal
G}_0\phi^*_{l,l}\phi_{l,l\pm1}+c.c\right)\nonumber
\end{eqnarray}
As usual in 1D systems, the spin and charge sectors in ${\cal
A}_{eff}$ are separated.

The last term in (\ref{a2}) may be represented as a loop
expansion. Two first diagrams are shown in Fig.2.
\begin{figure}
\includegraphics[width=0.3\textwidth]{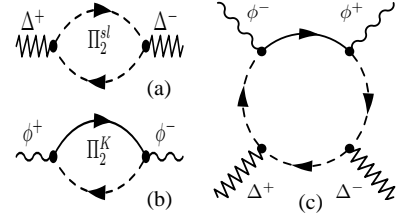}
\caption{\label{fig:f2a}Loop expansion for non-local action
(\ref{a2}) Solid and dashed lines in $\Pi_2$ and $\Pi_4$ stand for
electron and spinon propagators, respectively.}
\end{figure}
To calculate the diagrams, we use the non-local spinon GF ${\cal
D}^0_{j,j+r}(\omega_n)=({\cal D}_{loc}-\bar \Delta)^{-1}$ with
cosine-like dispersion $$ {\cal
D}^0_{j,j+r}(\omega_n)=\frac{\exp\left(-|r|\left[\ln\left(\frac{\omega_n-\sqrt{\omega_n^2+\bar\Delta^2}}{\bar\Delta}\right)
+i\frac{\pi}{2}\right]\right)}{i\sqrt{\omega_n^2+\bar\Delta^2}}.
\label{d2}
$$
Here {$r$ numerates sites in the chain, $\omega_n=2\pi T(n+1/4)$
on the imaginary axis \cite{pop}.
 ${\cal
D}^0$ is characterized by two branch cuts at $[\bar\Delta,
+\infty)$ and $(-\infty,-\bar\Delta]$. In the limit $\bar\Delta\ll
\pi T$ it rapidly falls down with growing $|r|$ as ${\cal
D}^0_{j,j+r}(\omega_n)\sim
-\bar\Delta^{|r|}/(-i\omega_n)^{|r|+1}.$ Thus the main
contribution comes from ${\cal
D}^0_{j,j}(\omega_n)=-i/\sqrt{\omega_n^2+\bar\Delta^2}$ and ${\cal
D}^0_{j,j\pm
1}(\omega_n)=(\omega_n/\sqrt{\omega_n^2+\bar\Delta^2}-1)/\bar\Delta$.

The polynomial effective action after the loop expansion is
performed is given by
\begin{eqnarray}
&&{\cal A}_{eff}= \sum_{\langle
jj'\rangle\omega_n}\left[\frac{|\Delta_{jj'}(\omega_n)|^2}
{I}-\Pi^{sl}_2|\Delta_{jj'}(\omega_n)-\bar\Delta|^2\right]\nonumber\\
&&+\sum_{jj'l,\omega_n}\left(\frac{1}{\tilde
J_{jl}}-\Pi^K_2+\Pi^K_4|\Delta_{jj'}(\omega_n)|^2\right)|\phi_{lj}(\omega_n)|^2
\nonumber\\
&& +Tr\log({\cal G}_0^{-1})+Tr\log[({\cal D}^0)^{-1}]+O(|\phi|^4).
\label{a3}
\end{eqnarray}
The polarization loops $\Pi_2$ and $\Pi_4$ are shown in Fig.2. The
action (\ref{a3}) is gauge invariant in accordance with Elitzur
theorem \cite{lark}, and spin and charge modes are separated both
in the 3D lattice and in the Fock space.

To estimate $\bar \Delta$, we refer to the properties of spin
chains with AF coupling \cite{schulz}. The quasi-long-range-order
in these chains may be treated in terms of boson excitations in
Luttinger liquid (LL) or fermion pairs in spin liquid. The spin
susceptibility of a chain, $\langle \Delta^+\Delta^-
\rangle_{\omega=0}\sim\bar \Delta^2$, acquires Pauli form at $T^*
\sim 8J^2/E_F$ \cite{egg}, so we assume  $\bar \Delta \sim T^*$ in
our estimates. This means that even in the critical region of
Doniach's diagram, $T_K \approx \tilde J^2/E_F,$ the spins are
"molten" into spin liquid at $T \sim T_K$, and the Kondo screening
is irrelevant at low $T$.

 Evaluation of $\Pi_2, \Pi_4$ in the limit $\pi$$T$$\gg$$\bar\Delta $
 gives  $\Pi^K_2$$\sim $$\rho_0$$\ln(\bar\Delta/T)$ and
 $\Pi^K_4$$\sim$$ \rho_0/\bar \Delta^2$. This
leads to reduction of indirect exchange, $\tilde I$$= $$I[1$$+$$
I/\bar \Delta $$\ln(\bar \Delta/T_K)]^{-1}.$ Since the Kondo
screening is quenched at $T^*$$>$$T_K$, we remain in a weak Kondo
screening regime even in the LL limit at $T$$\to $$0$. The
screening results in reduction of LL sound velocity, $\hbar
v$$=$$\tilde I a_z.$ As to the in-plane charge excitations, the
formation of Kondo clouds is quenched at $T$$\gg $$T_K$, so
instead of coherent Fermi liquid regime, $\langle \phi^+
\phi^-\rangle_{\omega\to 0}$ behaves as a relaxation mode $\sim
[-i\omega/\Gamma +\alpha q^2 + \ln (\bar \Delta/T_K)]^{-1},$ where
$\Gamma,\alpha$ are numerical constants.

These features of two-component electron/spin liquid manifest
themselves in thermodynamics. The logarithmic corrections $\sim
\ln^{-1}(T^*/T)$ are expected in low-$T$ Pauli-like susceptibility
of isotropic spin chains, whereas the logarithmic corrections to
the susceptibility of charged layers are quenched as $\ln  \bar
\Delta /T_K$. The overdamped relaxation mode should be seen as a
quasielastic peak in $\chi_0$. The 1D spinons contribute to the
linear-$T$ term in specific heat thus mimicking the heavy-fermion
behavior, while the contribution of Kondo clouds is frozen at low
$T$.
\begin{figure}
\includegraphics[width=0.21\textwidth]{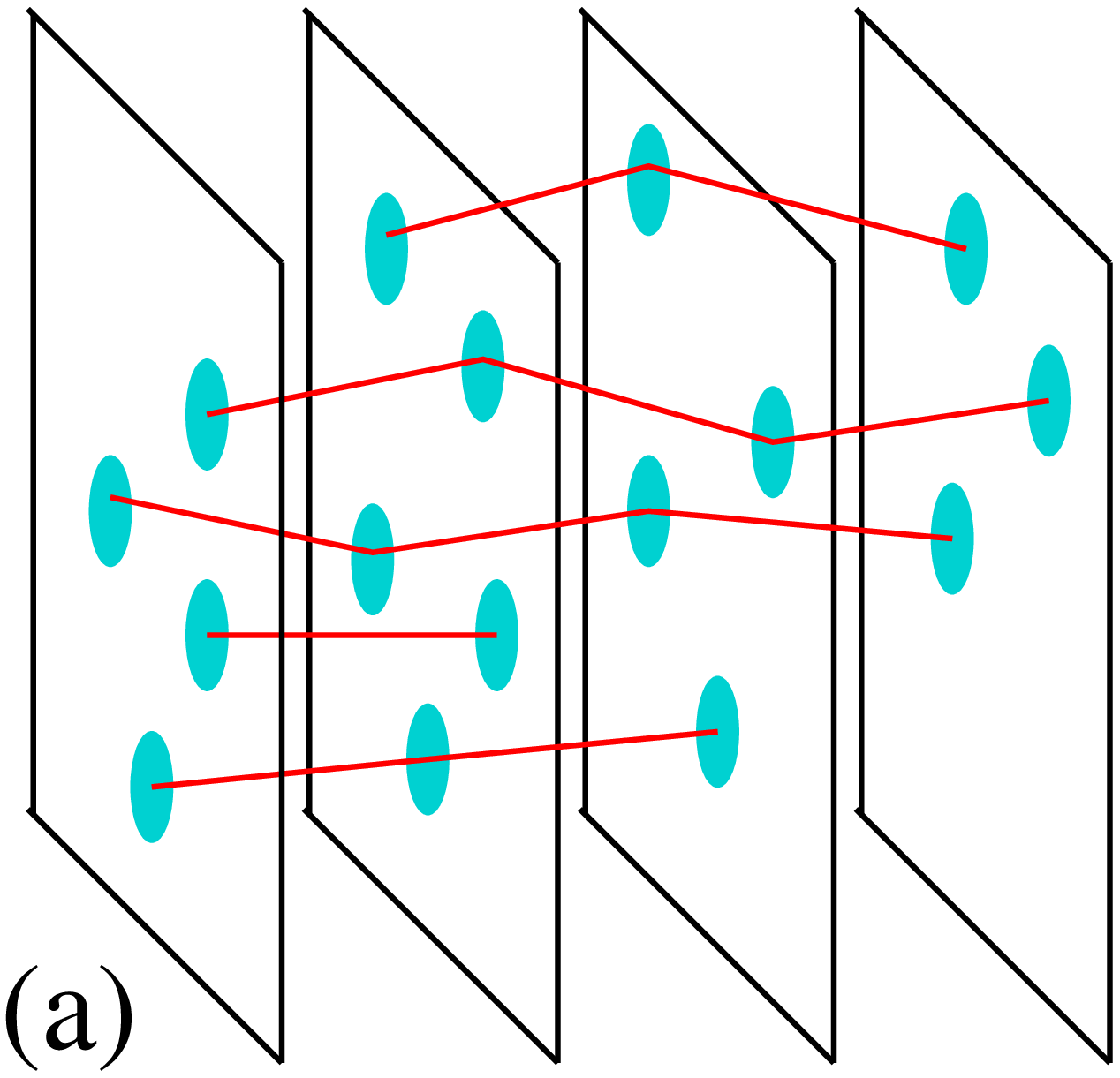}\;\;\;\;\;\;
\includegraphics[width=0.155\textwidth]{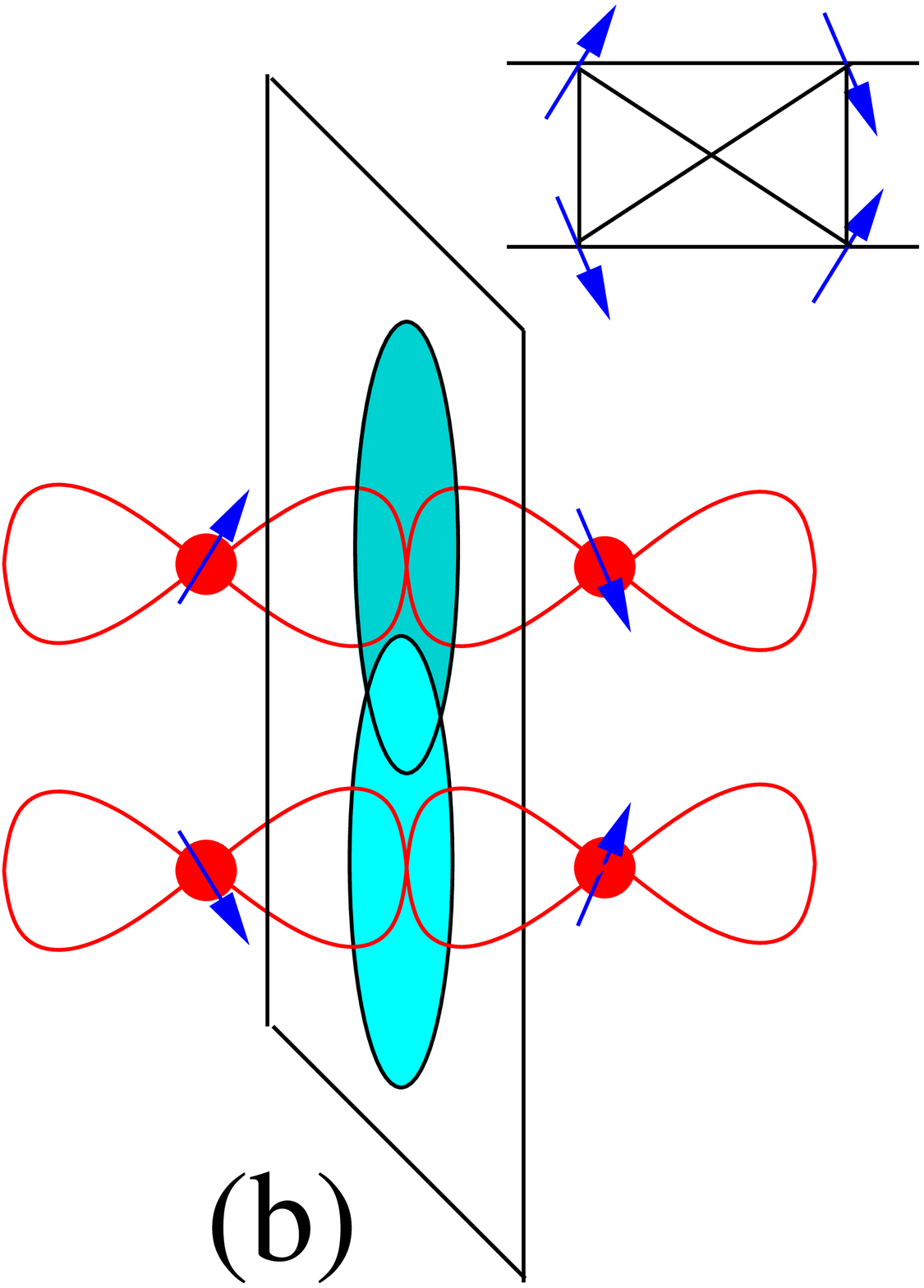}
\caption{(a) Disordered anisotropic Kondo lattices ; (b) Formation
of spin ladder from interacting chains. \label{fig:f4a}}
\end{figure}

In real anisotropic crystals one may expect formation of distorted
and dangling chains (Fig. \ref{fig:f4a}a) instead of an "ideal"
lattice (Fig. 1). Distortion means shift of two neighboring Kondo
"shadows" in a stack. This effect may be modelled by a random
overlap factor $w_j$ in RKKY integrals, $I_j=w_j I$. The dangling
bond effect means $w_j=0$. Bond disorder may be treated in terms
of random AFM chains \cite{fish}. According to this theory the
disorder results in transformation of singlet RVB liquid into a
random-singlet RVB phase  with arbitrarily long singlet bonds. In
 chains with broken bonds the gaps arise due to the finite length
effect, so the short chain segments does not contribute into the
low-T thermodynamics. With increasing impurity concentration, the
Kondo clouds begin to overlap and two-leg ladders with diagonal
bonds arise along with isolated chains (Fig. \ref{fig:f4a}b). The
Nozieres' exhaustion is still not actual for these clusters. With
further increase of the concentration of magnetic sites, the
Doniach's problem restores in its full glory. In case of FM
coupling $I$, true long-range order emerges in spin chains, but
the Nozieres' exhaustion is still quenched.

One may point out the class of layered conducting/magnetic hybrid
molecular solids as an object for application of above theory.
These crystals are formed by alternating metallic cationic layers
and insulating magnetic anionic layers with radicals $\rm
[N(CN)_2]^-$ as building blocks and transition metal ions as
carriers of localized spins \cite{expt}. Organic cations with
magnetic ions in such systems form ordered stacks. The problem is
in preparing metallic layers with large enough Fermi surface to
make Kondo screening effective and to find insulating networks
with large enough distance between magnetic ions. It is worth
noting that the above mentioned dicyanamide radicals with Mn ions
form planar Kagome sublattice, thus being a promising object for
realization of fractionalized Fermi liquid scenario proposed in
\cite{Sent1}.

This work is supported by SFB-410 project, ISF grant, A. Einstein
Minerva Center and the Transnational Access program $\#$
RITA-CT-2003-506095 at Weizmann Institute of Sciences.

\end{document}